\documentclass[superscriptaddress,twocolumn,showpacs,amsmath,amssymb,floats]{revtex4}

\usepackage{graphicx}% Include figure files

\usepackage{dcolumn}% Align table columns on decimal point

\usepackage{bm}% bold math

\usepackage{here}% force figure placement
\newcommand{\bb}{\begin{equation}}
\newcommand{\en}{\end{equation}}

\begin{document}

\title{Fluctuations and Rheology in Active Bacterial Suspensions}

\author{D.T.N. Chen}
\affiliation{Department of Physics and Astronomy, University of
Pennsylvania, Philadelphia, PA 19104}
\author{A.W.C. Lau}
\affiliation{Department of Physics, Florida Atlantic University,
Boca Raton, FL 33431}
\author{L.A. Hough}
\affiliation{Department of Physics and Astronomy, University of
Pennsylvania, Philadelphia, PA 19104}
\author{M.F. Islam}
\affiliation{Department of Physics and Astronomy, University of
Pennsylvania, Philadelphia, PA 19104}
\author{M. Goulian}
\affiliation{Department of Physics and Astronomy, University of
Pennsylvania, Philadelphia, PA 19104}
\author{T.C. Lubensky}
\affiliation{Department of Physics and Astronomy, University of
Pennsylvania, Philadelphia, PA 19104}
\author{A.G. Yodh}
\affiliation{Department of Physics and Astronomy, University of
Pennsylvania, Philadelphia, PA 19104}

\date{\today}

\begin{abstract}

We probe non-equilibrium properties of an active bacterial bath
through measurements of correlations of passive tracer particles and
the response function of a driven, optically trapped tracer. These
measurements demonstrate violation of the fluctuation-dissipation
theorem and enable us to extract the power spectrum of the active
stress fluctuations. In some cases, we observe $1/\sqrt{\omega}$
scaling in the noise spectrum which we show can be derived from a
theoretical model incorporating coupled stress, orientation, and
concentration fluctuations of the bacteria.

\end{abstract}

% insert suggested PACS numbers in braces on next line


%       82.70.Dd        Colloids
%        5.40.-a        Fluctuation phenomena ... Brownian motion
%       83.60.Bc        Linear Viscoelasticity
%       87.17.Jj        Cell locomotion; chemotaxis and related directed motion
%       87.18.Hf        Spatiotemporal pattern formation in cellular populations

\maketitle

% body of paper here

{\em Active} complex fluid systems such as living cells
\cite{Bursac,Lau}, assemblies of motors and filaments \cite{Furst},
flocks of birds \cite{Toner}, and vibrated granular media
\cite{Narayan} differ from conventional equilibrium media in that
some of their components consume and dissipate energy, thereby
creating a state that is far from equilibrium. An understanding of
model active systems, even at a phenomenological level, provides
insight about fundamental non-equilibrium statistical physics and,
potentially, about the inner workings of biological systems.
Bacterial baths \cite{Berg,Wu,Soni,Tuval,gregoire} are attractive
model active systems because energy input is homogeneous, because
individual bacteria can be directly observed, and because critical
parameters such as density, activity, and swimming behavior
\cite{Berg} can be brought under experimental control. Indeed,
recent experiments have reported on a rich variety of
non-equilibrium phenomena in this system class including anomalous
diffusion \cite{Wu} and pattern formation \cite{Tuval,Park}, while
recent theories of self-propelled organisms predict ordered phases
such as ``flocks" \cite{Toner}, their instabilities \cite{Simha},
novel rheological effects \cite{Hatwalne}, and giant density
fluctuations \cite{Chate}.

In this Letter, we describe measurements of the fluctuations and
mechanical responses of an active bacterial suspension. In contrast
to previous work \cite{Tuval, Wu, Soni}, we concurrently measure the
one- and {\em two-point} correlation functions of embedded passive
tracer particles to assess material fluctuations over a wide range
of length scales \cite{Crocker}, and we independently measure the
macroscopic response function of the ``active" medium by displacing
a particle through it with an optical trap \cite{Hough}. Even at a
low volume fraction ($\phi \sim 10^{-3}$) of bacteria, fluctuations
in the medium are substantially greater than they are in the absence
of bacteria while rheological response is unchanged, implying a
strong violation of the fluctuation-dissipation theorem (FDT). The
mean-square displacements (MSDs) of tracer particles as a function
of time $\Delta t$, depend on swimming behavior. For wild-type
bacteria, the MSD extracted from two-point correlations scale
superdiffusively as $\Delta t^{3/2}$ for the time scale of our
experiments, and the stress power spectrum \cite{Lau} as a function
of frequency $\omega$ scales as $\phi/\sqrt{\omega}$. Existing
theories of active media \cite{Simha} predict long-time tails and
anomalous corrections to diffusion but not superdiffusion. We
propose a theoretical model, following Ref.\ \cite{Hatwalne}, that
accounts for our experimental observations.

\begin{figure}
\includegraphics[height=2.0in]{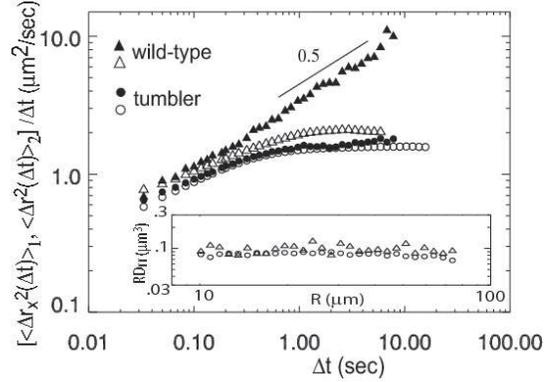}
\caption{1-pt (open symbols) and 2-pt (closed symbols) mean square
displacements divided by the lag time, $\Delta t$, for 2a = 2 $\mu$m
particles in bath of wild-type RP437 (triangles) and tumbling RP1616
(circles) bacteria at $\phi$ = .003. Inset: $RD_{rr}(R,\Delta t =
.067\,\mbox{sec}$) demonstrating $D_{rr} \sim 1/R$ for $R \geq 10 \,
\mu$m, and implying correlation length $\xi_U \leq 10 \, \mu m$.
\label{figure 1}}
\end{figure}

We used two strains of {\it E. coli}, a rod-shaped bacterium with
dimensions $3 \times 1 \,\mu \mbox{m}$, in these studies:  RP437,
the ``wild-type", which runs and tumbles \cite{Parkin} and RP1616,
the ``tumbler", which predominantly tumbles \cite{Alon}. Overnight
cultures were diluted 1/300 in Luria Broth (Difco) and grown at
$25\,^{\circ}\mbox{C}$ for 6 hrs. Subsequently, they were
centrifuged for 10 minutes at 5000 rpm and resuspended to the
desired concentration in a buffer comprised of 10 mM
K$_{2}$HPO$_{4}$, 0.1 mM EDTA, and 0.2 wt \% glucose (pH = 8.2),
which was added to maintain bacterial motility. We added a small
amount ($\phi_s = 10^{-4}$) of fluorescently labeled polystyrene
spheres (Duke Scientific) of radius $a$ to the bacterial suspension,
and to density match them, we added 15 wt \% sucrose to the
solution. To prevent bacterial adhesion, we prepared the chambers
from BSA coated glass slides and coverslips. We used parafilm
spacers to bring the thickness of the chambers to $\sim
240\,\mu\mbox{m}$; and we recorded quasi-2D image slices from the
middle of the 3D chamber. Samples were loaded into the chamber and
sealed with optical glue just prior to each run.

We quantified the fluctuations in the bacterial bath by computing
MSDs from the motions of embedded micron-sized tracers
\cite{Crocker}. The one-point displacement (MSD1) is defined by
$\langle \Delta {\bf r}^2(\Delta t) \rangle_1 = \sum_i \langle
\Delta r_i(t,\Delta t) \Delta r_i(t,\Delta t) \rangle$, where
$\Delta r_i(t,\Delta t) = r_i(t+\Delta t) - r_i(t)$ is the particle
displacement in the $i = (x, y, z)$ direction in time $\Delta t$,
and the brackets denote time and ensemble averaging. The {\em two-point}
displacement (MSD2) is defined as $\langle \Delta {\bf r}^2(\Delta
t)\rangle_2 = (2R/a) \, D_{rr}(R,\Delta t)$, where $D_{rr}(R,\Delta t)$
is the longitudinal component $D_{rr} (R , \Delta t) = D_{ij} R_i R_j/R^2 $
of the two-point tensor $D_{ij} ( R,\Delta t) = \langle \Delta r_i^{(1)}(t,\Delta
t) \Delta r_j^{(2)}(t, \Delta t) \rangle$, which measures correlations of
two distinct particles $(1,2)$ with an initial separation ${\bf R}$.  Over the
time scale of our experiments, ${\bf R}$ lies in the focus plane of our
microscope and its magnitude $R \equiv |{\bf R}|$ is greater than
that of individual particles' displacements.
The main advantage of two-point microrheology is that it provides a more
reliable measure of length scale dependent fluctuations in media
where the length scale of heterogeneities and tracer boundary
conditions are not {\it a priori} known \cite{Crocker, Lau}. Indeed, since
$D_{ij}(R, \Delta t)$ is ensemble averaged over tracer pairs with
$ R \gg a$, it reflects the dynamics of the medium on larger length scales than
the tracer size, permitting quantitative measurements even in the
presence of heterogeneities.  In general, MSD2 will equal MSD1 if
heterogeneities in the medium have length scales smaller than the
tracer size, otherwise they will differ in both magnitude
and functional form.

Typical MSD data are presented in Fig.\ \ref{figure 1}, which shows
that the one-point MSD in both bacterial strains displays a
crossover from superdiffusive behavior at short lag times ($\langle
\Delta r^2 \rangle_1 \sim \Delta t^{\alpha}, 1 < \alpha < 1.5 $) to
diffusive behavior ($\alpha = 1$) at long lag times. This
observation is similar to that of Ref.\ \cite{Wu}. However, our
two-point data for the wild-type, by contrast, exhibit a nearly
power-law superdiffusion ($\langle \Delta r^2 \rangle_2 \sim \Delta
t^{1.5}$) over 2.5 decades of observation time.  We also verified
that $D_{rr}(R,\Delta t) \sim 1/R$ (see inset of Fig.\ \ref{figure
1}), indicating that the bacterial bath, though an active medium,
can be treated on the separation scale $R$ as a coarse-grained
continuum  whose properties can be probed with two-point
microrheology \cite{Lau, Crocker}.  That MSDs exhibit superdiffusion
is suggestive of but not a proof of violation of FDT, which
requires an independent measurement of the rheological response
function.

\begin{figure}
\includegraphics[height=2.0in]{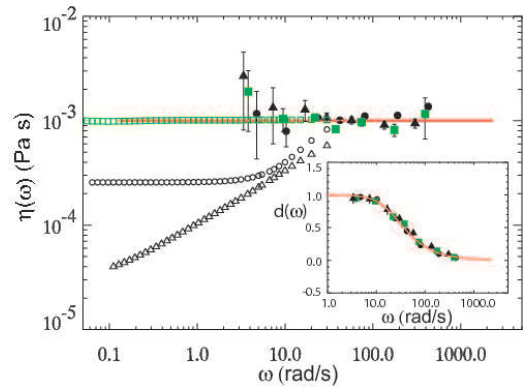}
\caption{(color online) Frequency dependent viscosity derived from
oscillating trap measurements for 4.0 $\mu$m sphere in water (solid
squares), the tumbler (solid circles), and the wild-type (solid
triangles) at $\phi$ = .003. Viscosities $\eta_2(\omega)$ derived
from the averaged two-point measurements using the generalized
Stokes-Einstein relation are plotted for the tumbler (open
circles), the wild-type (open triangles), and a bead in water (open
squares). Inset: Normalized Displacement of a 4.0 $\mu$m sphere in
the optical trap as a function of driving frequency for wild-type
(triangles), tumbler (circles), and water (squares). Line is a fit
to $d(\omega)$ (see text). \label{figure 2}}
\end{figure}

Response measurements were performed using an oscillating optical
tweezer setup similar to that of Ref.\ \cite{Hough}. Briefly, an
optical trap with typical trap stiffness of $\sim 1 \times 10^{-3}
\, \mbox{pN/nm}$ was formed by focusing an $\sim 100$ mW laser beam
($\lambda = 1054\, \mbox{nm}$) through a 1.3 NA oil immersion
objective (Zeiss). The trapping beam position was sinusoidally
scanned using a galvo-mirror at frequencies from 0.5 to 500 Hz. A
4.0 $\mu\mbox{m}$ PS sphere was trapped $\sim 6 \, \mu \mbox{m}$
from the coverslip. The position of the tracer was detected using
forward scattered light from a co-linearly aligned HeNe laser beam
focused onto a split photodiode (Hamamatsu S4204). The photodiode
signal was fed into a lock-in amplifier (Stanford Research Systems
530) along with the reference from the driving function generator
signal. The displacement and phase of the trapped particle output by
the lock-in amplifier were logged into a PC running LabView
(National Instruments).

The equation of motion for a particle of radius $a$ trapped in an
oscillating harmonic potential may be written as: $6 \pi \eta a
\dot{x} = -k\left [ x - A\cos(\omega t)\right ]$, where $\eta$ is
the viscosity of the medium, $k$ is the stiffness of the trap, and
$A$ is the driving amplitude.  Its steady state solution yields
the normalized displacement of the sphere in the trap: $d(\omega)
= \left \{ 1+ \left [\,6 \pi a \eta(\omega)  \omega/k\,\right
]^2\, \right \}^{-1/2}$.

The inset of Fig.\ \ref{figure 2} shows the raw normalized
displacement data for a particle in water and for a particle in an
active bacterial bath (RP1616). The solid line is a fit to
$d(\omega)$ with $\eta = 0.001\,\mbox{Pa $\cdot$ s}$, trap stiffness
$k = 8 \times 10^{-4}\,\mbox{pN/nm} $, and radius $a = 2.0\,\mu$m.
Both sets of experimental data agree with each other and with the
theoretical curve. From them, we extract the viscosity $\eta
(\omega)$ shown in the main graph of Fig.\ \ref{figure 2}. Clearly,
the presence of actively swimming bacteria at volume fraction
$10^{-3}$ does not modify the viscosity of the medium from that of
water, $\eta(\omega) = \eta_0 = 0.001\,\mbox{Pa $\cdot$ s}$. We
measured the same $\eta(\omega)$ using a capillary viscometer.

\begin{figure}[h]
\includegraphics[height=4.0in]{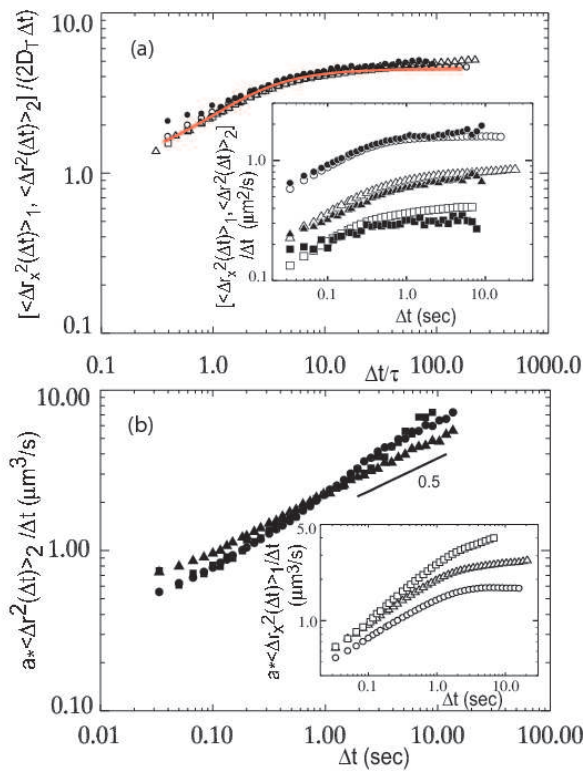}
\caption{(color online) (a) Collapsed 1-pt (open symbols) and radius
collapsed averaged 2-pt (closed symbols) MSDs for the tumblers at
$\phi$ = .003. The solid line is the master curve: $\gamma + (\,1 -
\gamma\,)(\,1- e^{-x}\,)/x$. Circles, triangles, and squares are for
particle diameters $2a = 2.0, 5.0,$ and $10.0$ $\mu$m, respectively.
Inset: Raw 1-pt (open symbols) and 2-pt (closed symbols) MSDs for
tumblers. (b) Radius rescaled 2-pt (closed symbols) MSDs for the
wild-types at $\phi$ = .003. Circles, triangles, and squares are for
particle diameters $2a = 2.0, 5.0,$ and $10.0$ $\mu$m, respectively.
Inset: Radius rescaled 1-pt MSD (open symbols) for the same $\phi$
and particle size data.} \label{figure 3}
\end{figure}

While recent theories of active systems predict novel enhancement in
the viscosity \cite{Hatwalne}, our experiments are well below the
concentration at which these effects are observable. Instead, our
results are consistent with the Einstein result for hard spheres:
$\eta = \eta_0 (1 + \frac{5}{2}\phi)$, namely, a negligible
modification in the viscosity for $\phi \sim 10^{-3}$. Moreover,
assuming for the moment the generalized Stokes-Einstein relation, we
can extract the (FDT consistent) response from the collapsed
two-point displacement (MSD2) \cite{Crocker}: $\eta_2(\omega) =
k_BT/3 \pi \omega^2 a \langle \Delta r^2(\omega) \rangle_2$, as
shown in Fig.\ \ref{figure 2}. The difference between $\eta
(\omega)$ and $\eta_2(\omega)$ explicitly indicates a strong
violation of FDT.

Next, to access the heterogeneity of the bacterial bath, we explored
the length-scale dependence of fluctuations by systematically
varying the size of the tracers at a fixed bacterial concentration.
The inset of Fig.\ \ref{figure 3}a shows MSDs obtained for spheres
in the tumbler bath. All samples and all tracer sizes exhibit a
crossover from superdiffusion to diffusion on similar timescales,
with an enhanced diffusion coefficient $D = \gamma D_T$, where
$\gamma = 4.3$ and $D_T = k_B T/(6 \pi \eta_0 a)$ is the equilibrium
coefficient. Moreover, MSD1 and MSD2 are nearly equal in magnitude
and functional form. Rescaling time by the crossover time $\tau$ and
the MSDs by $2D_T \Delta t$ collapses all the data onto a master
curve: $[ \langle \Delta r_x^2(\Delta t) \rangle_1 , \langle \Delta
r^2(\Delta t) \rangle_2 ]/(2 D_T \Delta t) = \gamma + (\,1 -
\gamma\,)(\,1- e^{-x}\,)/x$, where $x = \Delta t/\tau$. Figure
\ref{figure 3}a shows the collapsed MSD data along with the master
curve with $\tau = 0.1\,\mbox{s}$.

The MSDs for the wild-type are strikingly
different: the MSD1 exhibits a crossover dependent on tracer size,
while all of the MSD2 exhibit superdiffusion with nearly the same
exponent of $1.5$ over 2.5 decades of time, independent of the
tracer size. The trivial rescaling $a \langle \Delta r^2(\Delta t)
\rangle_2$ collapsed the respective MSD2 data [Fig.\ \ref{figure
3}b]. Under this rescaling, however, (and other simple scaling
functional forms as well) the wild-type MSD1 failed to collapse
[inset of Fig.\ \ref{figure 3}b], signaling the presence of
heterogeneity on the tracer length scale. The superdiffusive
exponent of the MSD1 approaches that of the two-point data ($\alpha
\sim 1.5$) as $a$ increases. This suggests that one-point
measurements are intrinsically ambiguous: the activity inferred
depends on the tracer size and boundary conditions \cite{Lau,Chen}.
Two-point measurements, in contrast, provide a more robust
characterization of the long-wavelength fluctuations of the medium
than one-point measurements.

\begin{figure}[h]
\includegraphics[height=2.5in]{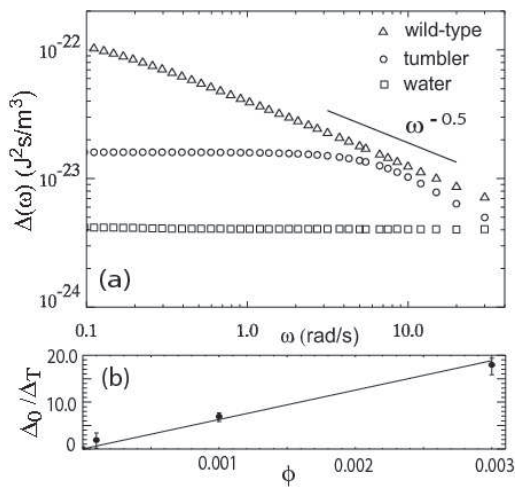}
\caption{ (a) The spectrum $\Delta (\omega)$ of active stress
fluctuations obtained from two-point microrheology and active
response measurements. The triangles are the wild-types, circles are
the tumblers (both $\phi = .003$), squares are water ($\phi = 0$).
(b) Linear dependence of the prefactor $\Delta_0$ in
$\Delta(\omega)$ on the volume fraction $\phi$ of the wild-type
bacteria; $\Delta_T \equiv 2\eta k_B T $. \label{figure 4}}
\end{figure}

We employ a recently developed phenomenological theoretical
framework for an active medium to interpret the experimental MSD
data \cite{Lau}. The bacterial activity gives rise to non-thermal
stress fluctuations whose power spectrum $\Delta(\omega)$ can
unambiguously be extracted from \textit{two-point} microrheology
data via
\begin{equation}
D_{rr}(R, \omega) = \frac{\Delta(\omega)}{6 \pi \omega^2 \,R
\,|\,\eta(\omega)|^2 }.
\label{drr}
\end{equation}
The results are exhibited in Fig.\ \ref{figure 4}a. For water, the
power spectrum is flat. For the tumblers, it is nearly Lorentzian,
flat at low frequencies with a knee at higher frequencies. For
wild-types, it exhibits power-law behavior, $\Delta(\omega)
\thicksim \omega^{-0.5}$, over 2.5 decades. In both cases,
$\Delta(\omega)$ is substantially greater than in a thermal system.
For the wild-type, the prefactor $\Delta_0$ of $\Delta(\omega)$
rises linearly with the bacterial concentration, as shown in Fig.\
\ref{figure 4}b.

We propose a simple model to account for the observed spectra of the
tumblers and the wild-types within the theoretical framework of
Ref.\ \cite{Hatwalne}. In the process of tumbling or swimming, each
bacterium contributes an additional active stress to the medium. It
has the form: $\sigma_{ij}^A \propto c({\bf x}, t) S_{ij}({\bf x},
t)$, where $c({\bf x}, t)$ is the concentration of the bacteria and
$S_{ij}$ is a force-dipole density generated by the active bacteria.
In wild types and tumblers, forces are directed, respectively, along
and perpendicular to the long-bacterial axes. Thus, in wild-types,
$S_{ij}$ is equal to the uniaxial nematic order parameter $Q_{ij}^U$
whereas in tumblers, it is equal to a biaxial order parameter
$Q_{ij}^B$. Active processes enhance stress fluctuations, and
assuming long-range isotropy, the active stress fluctuations can be
expressed as $\langle \,\sigma^A_{ij}({\bf q}, \omega )
\sigma^A_{kl}(- {\bf q}, - \omega )\,\rangle =
\Delta_{\sigma}(q,\omega) \left [ \delta_{ik} \delta_{jl} +
\delta_{il} \delta_{jk} - \frac{2}{3}\, \delta_{ij}\delta_{kl}
\right ]$ for both tumblers and wild-types. The power spectrum in
Eq.\ (\ref{drr}) is related to $\Delta_{\sigma}(q,\omega)$ by
$\Delta(\omega) = \Delta_{\sigma}({\bf q}=0, \omega)$.

The linearized equation for $Q_{ij}^A$ ($A = U,B$) is
$\partial_t Q_{ij}^A = - \tau_A^{-1}\left ( 1 - \xi_A^2 \nabla^2 \right )
Q_{ij}^A + s_{ij}$ , where $\tau_A$ is the relaxation time, $\xi_A$ the
correlation length of $Q_{ij}^A$, and $s_{ij}$ is a spatial-temporal white noise with
zero mean. Interactions among bacteria favor
long-range order in $Q_{ij}^U$ but not in $Q_{ij}^B$, implying that
$\tau_B \ll \tau_U$ and $\xi_B \ll \xi_U$. In both cases, the
concentration of bacteria obeys the continuity equation: $
\partial_t \delta c = - \nabla \cdot {\bf J}$ with $J_i = - D
\partial_i \delta c - \alpha_2 c_0
\partial_j Q_{ij}^A +  \delta {J}_i$, where $c_0$ is the average concentration,
$D$ is the diffusion constant, $\delta {J}_i$ is a random current,
and the second term stems from the nonequilibrium driving of mass
flow \cite{Toner}.  These equations lead after Gaussian decoupling
to $\Delta_{\sigma}( {\bf q}, \omega ) = \Delta_{\sigma} ( q
\xi_{\sigma A}, \omega )$ with $\xi_{\sigma A} \simeq \xi_A$. In
tumblers $\xi_B$ is very small, and $\Delta_{\sigma}(\omega)$ can be
replaced by a Lorentzian $\Delta (\omega)$ with characteristic time
$\tau_B \sim 0.1\,\mbox{s}$ (Fig.~\ref{figure 4}) \cite{andy} in
both MSD1 and MSD2 implying agreement between $\langle \Delta {\bf
r}^2 (\Delta t) \rangle_1$ and $\langle \Delta {\bf r}^2 (\Delta t)
\rangle_2$. In wild-types, $\Delta_{\sigma}({q}\xi_{U},\omega )$ can
be replaced by $\Delta (\omega)$ in MSD2 when $R> \xi_{U}$ [with
$\xi_{U} \le 10\,\mbox{$\mu$m}$ (see inset of Fig.~\ref{figure 1})],
but not in MSD1 when it probes lengths shorter than $\xi_{U}$,
implying different values for $\langle \Delta {\bf r}^2 (\Delta t)
\rangle_1$ and $\langle \Delta {\bf r}^2 (\Delta t) \rangle_2$
[Fig.~\ref{figure 1}]. Our calculations yield $\Delta(\omega) \sim
{c_0/ \sqrt{\omega}}$ for $\omega \tau_U > 1$ to lowest order in
$c_0$, in agreement with our experiments \cite{andy}.  This result
arises essentially from the concentration fluctuations in the active
stress, which were ignored in previous theories.  Note also that one
might expect that swimming bacteria have a tendency to develop
long-range {\em polar} rather than the nematic order of our model,
but we find that incipient polar order yields $\Delta(\omega) \sim
c_0^2\,\omega^{-3/2}$ in clear disagreement with our measurements
\cite{andy}.

In conclusion, using a combination of passive 2-pt microrheology
and active response measurements, we have demonstrated that the
macroscopic stress fluctuations depend sensitively on microscopic
swimming behavior of the bacteria. When contrasted with other
active systems such as living cells, our results suggest that
departures from equilibrium proceed via non-universal mechanisms.

We thank J.S. Parkinson for the bacterial strains used in the
experiment. We thank B.\ Hoffman, J.C.\ Crocker, and P.\ Collings
for stimulating discussions. This work has been partially supported
by the NSF MRSEC (DMR-05-20020) and NASA (NAG8-2172).

\end{document}